# Antiphase boundary in CH$_3$NH$_3$PbI$_3$ repels charge carriers while promotes fast ion migrations

Shulin Chen[1,2,#], Changwei Wu[3,#], Qiuyu Shang[4,#], Caili He[5], Wenke Zhou[6], Jinjin Zhao[5], Jingmin Zhang[1], Junlei Qi[7], Qing Zhang[4,*], Xiao Wang[3,*], Jiangyu Li[3,8,*], Peng Gao[1,9,10,11,*]

[1]Electron Microscopy Laboratory, School of Physics, Peking University, Beijing 100871, China

[2]KAUST Catalysis Center (KCC), Division of Physical Sciences and Engineering, King Abdullah University of Science and Technology (KAUST), Thuwal, 23955-6900, Saudi Arabia

[3]Shenzhen Key Laboratory of Nanobiomechanics, Shenzhen Institute of Advanced Technology, Chinese Academy of Sciences, Shenzhen 518055, China

[4]School of Materials Science and Engineering, Peking University, Beijing 100871, China

[5]School of Materials Science and Engineering, School of Mechanical Engineering, Shijiazhuang Tiedao University, Shijiazhuang, 050043, China

[6]Army Engineering University of PLA, State Key Laboratory for Disaster Prevention & Mitigation of Explosion & Impact, Nanjing, 210007, China

[7]State Key Laboratory of Advanced Welding and Joining, Harbin Institute of Technology, Harbin 150001, China

[8]Guangdong Provincial Key Laboratory of Functional Oxide Materials and Devices, Department of Materials Science and Engineering, Southern University of Science and Technology, Shenzhen 518055, China

[9]International Center for Quantum Materials, School of Physics, Peking University, Beijing 100871, China

[10]Collaborative Innovation Center of Quantum Matter, Beijing 100871, China

[11]Interdisciplinary Institute of Light-Element Quantum Materials and Research Center for Light-Element Advanced Materials, Peking University, Beijing 100871, China

[#]These authors contributed equally: Shulin Chen, Changwei Wu, Qiuyu Shang

*Correspondence and requests for materials should be addressed to p-gao@pku.edu.cn (P. Gao); lijy@sustech.edu.cn (J. Y. Li); xiao.wang@siat.ac.cn (X. Wang); q_zhang@pku.edu.cn (Q. Zhang)

**Abstract:** Defects in organic-inorganic hybrid perovskites (OIHPs) greatly influence their optoelectronic properties. Identification and better understanding of defects existing in OIHPs is an essential step towards fabricating high-performance perovskite solar cells. However, direct visualizing the defects is still a challenge for OIHPs due to their sensitivity during electron microscopy characterizations. Here, by using low dose scanning transmission electron microscopy techniques, we observe the common existence of antiphase boundary (APB) in $CH_3NH_3PbI_3$ ($MAPbI_3$), resolve its atomic structure, and correlate it to the electrical/ionic activities and structural instabilities. Such an APB is caused by the half-unit-cell shift of $[PbI_6]^{4-}$ octahedron along the [100]/[010] direction, leading to the transformation from corner-sharing $[PbI_6]^{4-}$ octahedron in bulk $MAPbI_3$ into edge-sharing ones at the APB. Based on the identified atomic-scale configuration, we further carry out density functional theory calculations and reveal that the APB in $MAPbI_3$ repels both electrons and holes while serves as a fast ion-migration channel, causing a rapid decomposition into $PbI_2$ that is detrimental to optoelectronic performance. These findings provide valuable insights into the relationships between structures and optoelectronic properties of OIHPs and suggest that controlling the APB is essential for their stability.

## 1. Introduction

Organic-inorganic hybrid perovskites (OIHPs) hold great promise for the next-generation solar cells because of their impressive power conversion efficiency (PCE) and facile cost-effective processing route[1-4]. During the synthesis of OIHPs, the comparably low temperature and fast nucleation and crystallization from solution inevitably cause unintentional point and planar defects[5]. The defects density ($10^{16}$-$10^{17}$ cm$^{-3}$) in a solution deposited $CH_3NH_3PbI_3$ (MAPbI$_3$) film is much higher than that in a single crystal MAPbI$_3$ ($10^{10}$-$10^{11}$ cm$^{-3}$)[6]. These defects greatly influence the electrical and ionic activities and are considered to be responsible for the hysteresis, charge trapping and scattering, and ion migration in OIHPs, further causing inferior performance and instability [6]. For example, the point defects such as cation antistites and Pb interstitials in OIHPs cause deep-level defects and nonradiative recombination centers[7], which trap charges and limit the photovoltaic performance. Li et al. reported that the PCE and the lifetime of the carrier are deteriorated with the increased density of twinning and stacking faults in perovskite solar cells (PSCs)[5]. Moreover, the defect density at the grain boundary (GB) is several orders of magnitude higher than that inside of the grain[8] while GB is generally considered as a shortcut for ion migration[9], leading to large hysteresis[10]. Moreover, some suggest that GB is electrically benign and facilitates charge separation and collection[11, 12] while others propose that GB plays as the nonradiative recombination center and deteriorates the device performance[8].

Besides the point defects, twinning, stacking faults, and GB, antiphase boundary (APB) also commonly exists in the PSCs. Indeed, previous studies reported APB usually presents unique electrical and ionic properties that are absent in the bulk[13-15]. In oxide perovskites, atomic-resolved transmission electron microscopy (TEM) images show that APB displays an antipolar phase in La$_{2/3}$Sr$_{1/3}$MnO$_3$ and induces a giant piezoelectric coefficient in NaNbO$_3$[16]. In transition metal dichalcogenides, APB acts as a faceted metallic wire to facilitate electron transport[17] while it repels both electrons and holes in all-inorganic perovskite[18]. Reducing the APB defects in GaInP$_2$ films significantly increases the minority carrier lifetime and eliminates rapid carrier recombination[19]. APB also has a great impact on ion migration in functional devices. Kaufman et al. revealed that APB migration is a fundamental diffusion mechanism in sodium layered oxide with quite low kinetic barriers[20]. Also, APB provides additional diffusion channels for lithium-ion migration in Li$_x$CoO$_2$[21]. Heisig et al. find that APB constitutes fast cation diffusion in SrTiO$_3$ memristive devices and decreases the diffusion barrier of Sr$^{2+}$ from 4.0 eV to 1.3 eV[22], and thus SrTiO$_3$ memristive devices with intentionally induced APB requires no forming steps[23]. Considering the great impact of APB on electrical and ionic activities, which are closely related to the optoelectronic performance of PSCs, it is necessary to identify the atomic structure of APB in OIHPs and reveal how it influences the optoelectronic properties.

So far, there have been few reports of the atomic structure of APB in OIHPs, let alone its impact on the electrical and ionic activity. This is mainly because APB features a half-unit-cell shift of registry with respect to two adjacent regions. Although the TEM proves to be one of the most powerful tools to study APB[24], OIHPs are extremely sensitive to electron beam illumination[25, 26], making it challenging to observe atomic-scale structures of APB by TEM. Recently, Rothmann et al. have successfully observed the atomic structure, boundary, and defects of CH(NH$_2$)$_2$PbI$_3$ (FAPbI$_3$) by low dose scanning TEM (STEM) techniques[27]. In this work, we adopted similar low-dose STEM techniques to resolve the atomic structure of APB in MAPbI$_3$ and then clarified its impact on electrical and ionic activities via density functional theory (DFT) calculations. Atomic-scale images show that APB is composed of the edge-shared [PbI$_6$]$^{4-}$ octahedron and prefers to propagate along the [100] and [010] directions. Based on such an atomic structure, the effect of APB in MAPbI$_3$ on the electrical/ionic activity is clarified by DFT calculations. We find that while APB in MAPbI$_3$ does not introduce any deep-level defects and repels both electrons and holes, the diffusion barriers of CH$_3$NH$_3^+$ (MA$^+$), Pb$^{2+}$, and I$^-$ are lowered at the APB compared to that in the bulk MAPbI$_3$. These suggest that APB provides a fast ion-migration channel, facilitating a more facile decomposition of MAPbI$_3$ into PbI$_2$. These findings provide atomic-scale insights into the structure of APB in MAPbI$_3$ and clarify the influence of APB on electrical/ionic activity, which enhances our understanding of the correlations between structures and optoelectronic properties.

## 2. Results and discussion

Nanocrystal and polycrystal MAPbI$_3$ are chosen to investigate the atomic structures of the defects in MAPbI$_3$. Nanocrystal MAPbI$_3$ is about 10-20 nm large with good crystallinity (Figure S1). Polycrystalline MAPbI$_3$ thin film was directly grown on the ultrathin carbon-coated transmission electron microscopy (TEM) copper grids with each domain size around 100-300 nm (Figure S2). Since MAPbI$_3$ is sensitive to the electron beam[28-30], low-dose imaging techniques including direct-detection electron-counting (DDEC) camera and low-dose scanning transmission electron microscopy (STEM) were used to obtain the atomic structures of MAPbI$_3$ (Figure 1). Figure 1a shows high resolution TEM (HRTEM) image acquired at a dose of 28.2 e Å$^{-2}$ by DDEC camera from nanocrystal MAPbI$_3$. Yet there are many MA$^+$ vacancies as highlighted by the yellow squares in Figure 1a under electron beam illumination[31], thus the corresponding fast Fourier transform (FFT) pattern shows superstructure spots (Figure 1b). In contrast, the acquired STEM image at 96.4 e Å$^{-2}$ from polycrystal MAPbI$_3$ shows more visible atomic columns, despite that the obtained structure might suffer from larger damage and more MA$^+$ vacancies are formed (Figure 1c) with higher-order diffraction spots lost (Figure 1d). With the increased electron dose, MAPbI$_3$ gradually decomposes into PbI$_2$ within 964.0 e Å$^{-2}$ (Figure S3). For both of these two imaging techniques, MA$^+$ vacancies are inevitably generated even at a relatively low dose that is necessary for decent atomic structure visualization. The contrast of TEM images highly depends on the thickness of the sample and the imaging defocus and thus is less reliable to identify specific atomic columns, while the contrast of STEM image is easy to interpret and sensitive to the atomic number (Z). Accordingly, we adopt STEM techniques to characterize the atomic structures of the defects in MAPbI$_3$.

Figure 2a is a STEM image of the MAPbI$_3$ along the [001] direction. Some boundaries are indicated by the white lines. These boundaries prefer to lie along [100] and [010] directions. Since the contrast of STEM images is sensitive to Z, each type of the atomic columns can be identified as indicated in magnified Figure 2b. Note that MA$^+$ vacancies are formed due to the beam damage while the same structure without MA$^+$ is verified to be unstable (Figure S4), thus the atomic model of the boundary before forming MA$^+$ vacancies is proposed in Figure 2c,d. The structure transition from pristine MAPbI$_3$ (Figure 2c) to the boundary structure (Figure 2d) can be achieved as follows: the right region with green octahedrons shifts half of the unit cell along a/b direction, accompanied by the corner-sharing [PbI$_6$]$^{4-}$ octahedron transformed into edge-shared one. Indeed, this is a typical feature of the APB. Since electron beam usually leads to the formation of vacancies and local ion migration[25, 27, 31], such a collective shift of atomic columns of the APB defect should be the pristine feature of MAPbI$_3$ rather than induced by the electron beam illumination. The DFT-optimized atomic structure of APB is shown in Figure S5 and the formation energy of such APB is calculated to be 0.8 eV per unit, indicating APB is easy to form during the synthesis process. More representative APB defects are shown in Figure S6. In addition, similar APB structures have also been observed in all-inorganic perovskite (CsPbBr$_3$) as shown in Figure S7, suggesting such an APB defect is general in OIHPs and its all-inorganic counterpart. Note that such an APB structure is a 90° boundary, which is difficult to identify by electron diffraction or FFT patterns without atomic-scale imaging. Indeed, most previous electron microscopy studies failed to observe them, without which the effect on the material performance is impossible to establish.

Having obtained the atomic structure of APB, we are now ready to investigate how such an APB influences the electrical properties by DFT calculations. Figure 3a shows the density of state (DOS) of MAPbI$_3$ and APB. It is observed that the APB does not introduce any deep-level defects within the bandgap, which usually prevent charge transport and facilitate the nonradiative recombination. To reveal its effect on the electron and hole transport, we further examined the band diagram across the APB. Figure 3b presents a layer-by-layer projection of the DOS (LDOS) across the APB. A large bandgap offset can be observed across the APB. Specifically, the conduction band minimum (CBM) offsets +31 meV while the valence band maximum (VBM) offsets -41 meV at the APB, thus APB features a type-I band alignment, which efficiently repels both electrons and holes[32]. This result is consistent with the charge density of the CBM and VBM of APB as shown in Figure 3c-f. It is observed that the evenly-distributed charge density of CBM and VBM in the bulk MAPbI$_3$ decreases at the APB since the positive offset of CBM repels away electrons from APB and the negative offset of VBM also drives the holes away from it.

It is also desirable to clarify the influence of APB on ion migration, which is significant for PSCs. Previous studies show that MA$^+$ and I$^-$ are easy to migrate within MAPbI$_3$ while the diffusion barrier of Pb$^{2+}$ is higher[33], which can be induced by high temperatures[34]. By DFT calculations, we have compared the diffusion barrier of MA$^+$, Pb$^{2+}$, and I$^-$ at the APB to that in the bulk as shown in Figure 4. Figure 4a and Figure 4b show the schematic diagram of vacancy-mediated migrating pathways of these ions in the bulk MAPbI$_3$ and at the APB. Specifically, MA$^+$ diffuses to the neighboring vacant A-site while Pb$^{2+}$ migrates along the diagonal of the (110) plane through Pb$^{2+}$ vacancy and I$^-$ migrates along an edge of the octahedron, as illustrated in Figure S8. Figure 4c and Table S1 present the diffusion barrier of $MA^+$, $Pb^{2+}$, and $I^-$ along the corresponding pathways. The diffusion barrier in the bulk $MAPbI_3$ for $MA^+$, $Pb^{2+}$, and $I^-$ is 0.98, 2.37, and 0.55 eV, similar to the reported values[33] while the diffusion barrier of $MA^+$, $Pb^{2+}$, and $I^-$ decreases to 0.77, 1.38 and 0.43 eV at the APB. This suggests $MA^+$, $Pb^{2+}$, and $I^-$ are much easier to migrate along with the APB, which serves as a fast ion-diffusion channel and likely causes a more facile decomposition of $MAPbI_3$. As shown in Figure S9, $MAPbI_3$ with APB structures decomposes into $PbI_2$ within 385.6 e $Å^{-2}$, which is much lower than that of the bulk $MAPbI_3$ (964.0 e $Å^{-2}$) under the same electron dose rate.

### 3. Discussion and summary

The intrinsic optoelectronic properties of OIHPs are greatly influenced by the defects within the crystal[6, 35]. To fabricate high-performance PSCs, it is necessary to enhance the understanding of defects in OIHPs. The frequently-used techniques to characterize the defects like steady-state photoluminescence[36], space charge limited current[37], and thermally simulated current[38] can provide useful information about defects, but they are unable to identify the specific types of defects as well as their atomic structures. A previous study has observed the APB in $FAPbI_3$, though the elaborate atomic-scale configuration is still unknown[27]. By using low dose STEM techniques, we directly observed the existence of APB in $MAPbI_3$ and resolved its atomic structure. Based on the identified atomic-scale configuration of APB, we further clarified its influence on the electrical and ionic activities of OIHPs with the assistance of DFT calculations, resulting in an improved understanding on the relationship between defect structures and properties.

In traditional semiconductors, planar defects usually introduce deep-level defects within the bandgap, hindering the charges transport and facilitating the nonradiative recombination[39]. In contrast, our work reveals that the planar defect of APB in $MAPbI_3$ does not introduce any deep-level defects within the bandgap. Moreover, such APB repels the electrons and holes and features as a type-I band alignment. It has been reported that such type-I band alignment at the grain boundaries on the surface of OIHPs can effectively repel carriers and return them to the inside of grain, thus decreasing the carrier loss and facilitating an improved optoelectronic performance[32, 40]. The APB observed in our work is mainly inside the grain, and thus delicate engineering is necessary to control its distribution.

Furthermore, ion migration is regarded as one of the most important issues in PSCs, responsible for phase segregation, J-V hysteresis, and device degradation[41]. Such ion migration can be intrinsic due to the low migration energy. Our work finds that the APB provides additional ion diffusion channels and the diffusion barrier of $MA^+$, $Pb^{2+}$, and $I^-$ decreases by 21.8%, 41.8%, and 20.4% respectively at the APB. Such a low diffusion barrier induces easier ion migration and greatly increases the chemical activity of $MAPbI_3$, leading to more facile structure degradation (Figure S9) that destroys long-term operational stabilities[42]. In particular, the diffusion barrier of $Pb^{2+}$ decreases from 2.37 to 1.38 eV at the APB, making it easier to form $Pb^{2+}$ interstitials and $Pb^{2+}$-related antistites, both of which can create deep-level defect traps as recombination centers[7] and are detrimental to efficient charge extractions. This suggests efficient control and engineering of defects are highly desirable for high-performance PSCs. For example, reducing the density of the twin boundaries in $MA_{1-x}FA_xPbI_3$ via defect-engineering[5] and minimizing hydrogen vacancies[43] enable a much-improved performance of PSCs.

In summary, by using low dose STEM techniques, we have successfully observed the existence of APB in OIHPs, revealed its atomic structure, and further clarified its impact on electronic structure, ion migration, and structure instabilities. Atomic-resolution STEM images show that the APB consists of edge-sharing $[PbI_6]^{4-}$ octahedron and lies along the [100] and [010] directions. Further DFT calculations based on the identified atomic-scale configuration show that the APB repels both electrons and holes and facilitates fast diffusion of MA$^+$, Pb$^{2+}$, and I$^-$. The fast ion diffusion at the APB further leads to a quick decomposition into PbI$_2$. These findings enhance a better understanding of the relationships between structures and optoelectronic properties of OIHPs and suggest that controlling the APB is essential for their stability.

## 4. Experimental Section

**MAPbI$_3$ synthesis.** MAPbI$_3$ nanocrystals were bought from Xiamen Luman Technology Co., Ltd. MAPbI$_3$ films were grown directly on the ultrathin carbon-coated copper transmission electron microscopy (TEM) grids, as previously reported[44]. Specifically, the precursor solution was prepared by mixing 99.5% pure methlammonium iodide (MAI) and 99.999% lead iodide (PbI$_2$) in dimethylformamide to get a 45 wt. % solution. Then the obtained precursor solution was deposited on ultrathin carbon-coated copper grids (300 mesh) by spin coating at 6,000 r.p.m. for 70 s. During this process, 50 μL chlorobenzene was dropped on the spinning substrate after 30 s, followed by annealing at 100 °C for 10 minutes. Thus the MAPbI$_3$ film can be obtained[45].

**CsPbBr$_3$ synthesis.** CsBr (0.4 mmol) and PbBr$_2$ (0.4 mmol) were dissolved in dimethylformamide (10 mL). 1 mL oleic acid and 0.5 mL oleylamine were added into the precursor solution. After, 1 mL precursor solution was fast added into 10 mL toluene with strong stirring. Then 1 mL solution was mixed with 4 mL methyl acetate, and centrifuged at 8,000 r.p.m. for 4 minutes, followed by dissolving into 1 mL toluene to get CsPbBr$_3$ crystals[46].

**Characterization.** The selected area electron diffraction patterns and STEM images were conducted at an aberration-corrected FEI (Titan Cubed Themis G2) operated at 300 kV. SAED images were obtained at 1 e Å$^{-2}$ s$^{-1}$. STEM images were acquired at a current of 1 pA, a convergence semi-angle of 21.4 mrad, and a collection semi-angle snap in the range of 25–153 mrad, which allows the efficient imaging of low-Z elements[47]. The corresponding dwell time is 0.5 μs and the pixel size is 18 pm. The dose rate at STEM mode is estimated by dividing the screen current by the area of the raster[48]. To reduce the electron beam damage, the spherical aberration and focus were adjusted away from imaged areas. Each image was obtained without adjusting the zone axis to decrease the beam damage.

HRTEM images were acquired by DDEC camera using electron-counting mode with the dose fractionation function. The drift was corrected by DigitalMicrograph software. The original image stack contains 40 subframes in 4 s. Atomistic models were constructed by Vesta software.

**Density functional theory calculation**. Our first-principles calculations were carried out within the framework of DFT as implemented in the Vienna ab initio simulation package code[49, 50]. The electron-ion interactions were described by the projector augmented-wave method[51]. The electron exchange-correlation was treated by a generalized gradient approximation with Perdew-Bruke-Ernzerhof functional[52]. The kinetic cutoff energy was set as 500 eV for the Kohn-Sham orbitals being expanded on the plane-wave basis. The supercell size of APB was repeated periodically along the [100] direction. The atomic positions and lattice constants were fully optimized with a conjugate gradient algorithm until the Hellman-Feynman force on each atom is less than 0.01 eV/Å[53]. The Monkhorst-Pack k- point meshes were sampled as 9×9×7 and 3×9×7 for the MAPbI$_3$ and APB, respectively[54]. The minimum energy pathways of ions migration were determined through the climbing image nudged elastic band method[55] based on the interatomic forces and total energies acquired from DFT calculations. We performed the Ab initio molecular dynamic (AIMD) simulation in a canonical ensemble. The Brillouin zone was sampled at the Γ point and the time step of the AIMD simulation is 1 fs.

**Supporting Information**

Supporting Information is available from the Wiley Online Library or the author.

**Acknowledgements**

This research was financially supported by the National Natural Science Foundation of China (11974023, 52021006, 11772207, 22003074, 52072006), Youth Innovation Promotion Association CAS, the Key R&D Program of Guangdong Province (2018B030327001, 2018B010109009), the Guangdong Provincial Key Laboratory Program from the Department of Science and Technology of Guangdong Province (2021B1212040001), Guangdong Basic and Applied Basic Research Foundation (2020A1515110580), Natural Science Foundation of Hebei Province for distinguished young scholar (A2019210204) and China Postdoctoral Science Foundation (2021M703391). The authors gratefully acknowledge the Electron Microscopy Laboratory at Peking University for the use of electron microscopes and the support of the Center for Computational Science and Engineering at Southern University of Science and Technology.

**Conflict of Interest**

The authors declare no conflict of interest.

**Author contributions**

P. Gao, J.Y. Li, and S.L. Chen conceived and supervised the project. S.L. Chen carried out TEM experiments and analysed experimental data with the direction of P. Gao and help from J.M. Zhang; C.W. Wu performed the calculations under the guidance of X. Wang. Q.Y. Shang synthesized MAPbI$_3$ thin films under the direction of Q. Zhang. C.L. He prepared the CsPbBr$_3$ materials under the guidance of J.J. Zhao. W.K. Zhou and J.L. Qi provided additional specimen. S.L. Chen, J.Y. Li, and P. Gao wrote the manuscript and all authors participated in the revision.

**Data Availability Statement**

The data that support the findings of this study are available from the corresponding author upon reasonable request.

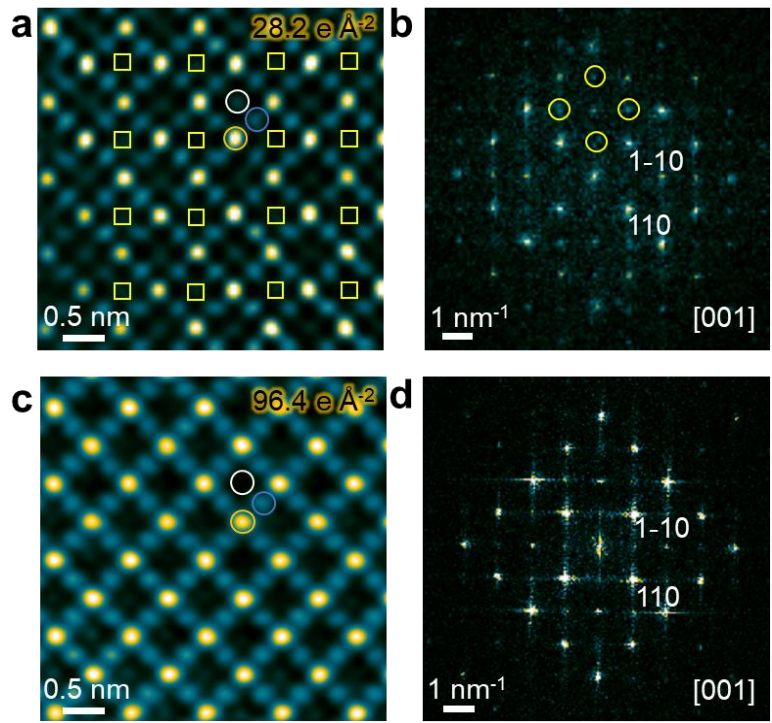

**Figure 1 Low-dose imaging of the atomic structure of MAPbI$_3$. a** HRTEM image of nanocrystal MAPbI$_3$ at a dose of 28.2 e Å$^{-2}$ by using a direct-detection electron-counting camera. The yellow circles show the Pb$^{2+}$/I$^-$ columns, the purple circles show pure I$^-$ columns and the white circles show the MA$^+$ columns. The yellow squares indicate the MA$^+$ vacancies. **b** The corresponding FFT pattern along the [001] direction. The yellow circles mark the superstructure diffraction spots. **c** STEM image of MAPbI$_3$ film directly grown on ultrathin carbon-coated TEM grids at a dose of 96.4 e Å$^{-2}$. The yellow, purple, and white circles represent Pb$^{2+}$/I$^-$, I$^-$, and MA$^+$ columns. **d** The corresponding FFT pattern along the [001] direction.

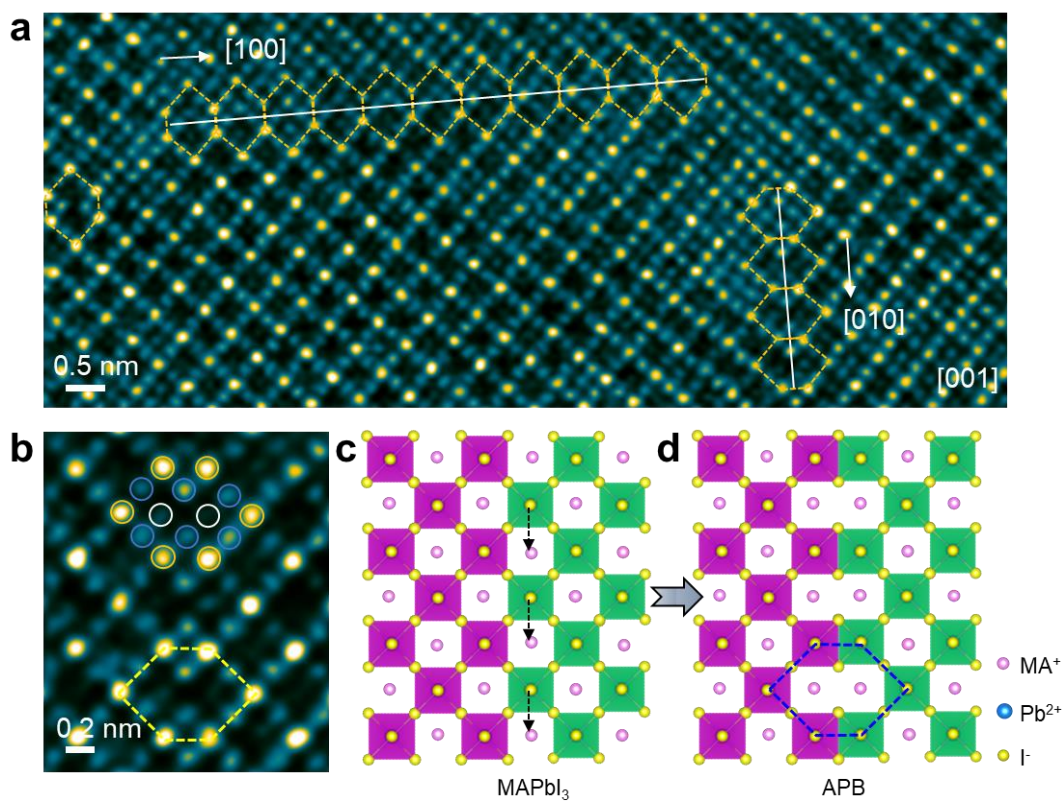

**Figure 2 Atomic structure of antiphase boundary in MAPbI$_3$. a** Atomic structure of APB along the [001] direction. The white lines highlight that APB prefers to lie along the [100] and [010] directions. **b** Enlarged view of the atomic structure of the APB. The yellow, purple, and white circles indicate Pb$^{2+}$/I$^-$, I$^-$, and MA$^+$ columns, respectively. **c** Atomic model of MAPbI$_3$. The [PbI$_6$]$^{4-}$ octahedrons are highlighted by the purple (left region) and green color (right region). **d** Atomic model of APB in MAPbI$_3$. After green octahedrons in MAPbI$_3$ shifts half of the unit cell along a/b direction, as the black arrows indicate, the structure transforms into APB. The corner-sharing octahedrons become edge-shared ones. The blue hexagon in **d** corresponds to the yellow one in **a** and **b**. Light blue, purple, and yellow balls indicate MA$^+$, Pb$^{2+}$, and I$^-$ respectively.

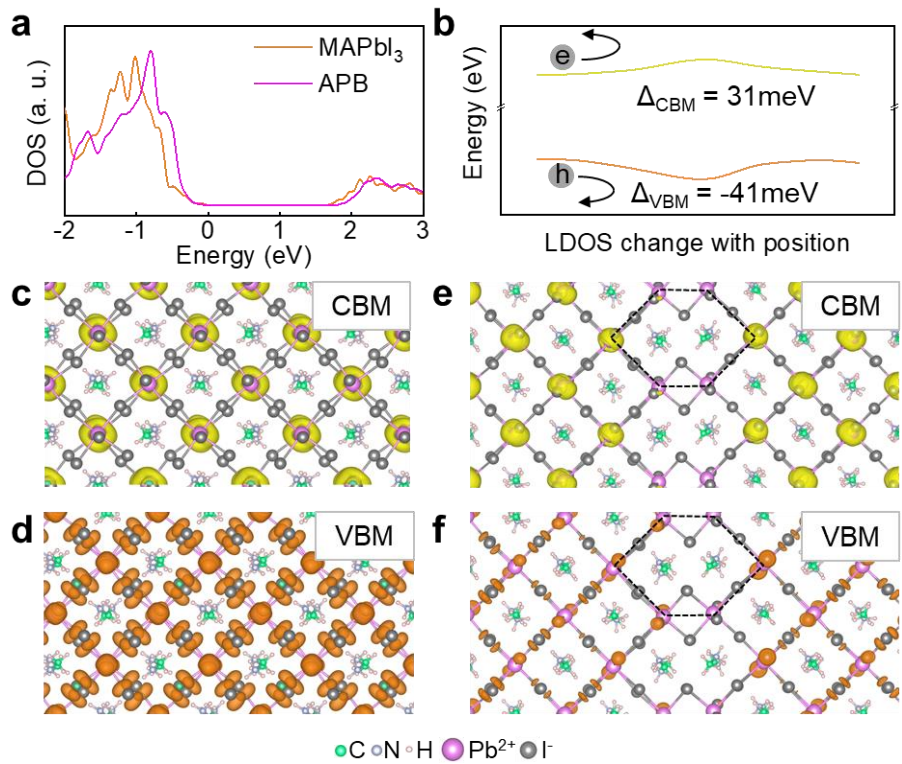

**Figure 3 The effect of APB on electrical properties. a** The DOS of the bulk MAPbI$_3$ and APB defect. **b** LDOS and band diagram of APB with a positive offset (31 meV) of the conduction band and a negative offset (-41 meV) of the valance band, thus repelling both electrons and holes. **c**, **d** Charge density of CBM and VBM of the bulk MAPbI$_3$. **e**, **f** Charge density of CBM and VBM of the APB defect.

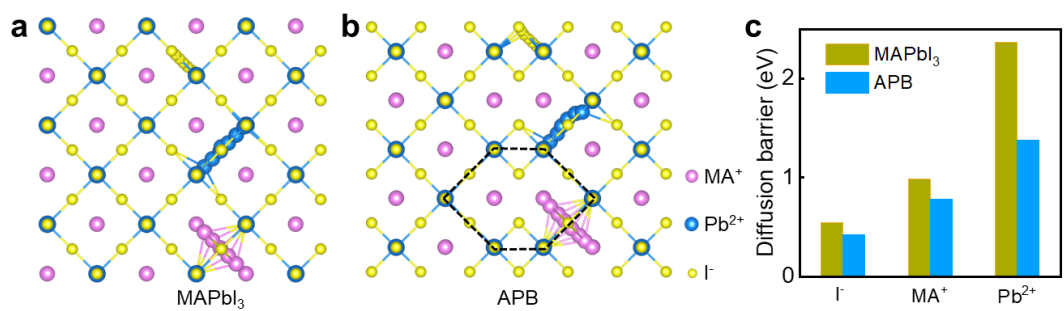

**Figure 4 The effect of APB on ion migration. a, b** Schematic diagram to illustrate the migration pathway of MA$^+$, Pb$^{2+}$, and I$^-$ in the bulk MAPbI$_3$ and at the APB. Light blue, purple, and yellow balls indicate MA$^+$, Pb$^{2+}$, and I$^-$ respectively. The specific pathway in the optimized structure can be found in Figure S8. **c** The diffusion barrier of MA$^+$, Pb$^{2+}$, and I$^-$ along the corresponding diffusion pathways in **a** and **b** at the APB and in the bulk MAPbI$_3$. The diffusion barrier has also been listed in Table S1.